\documentclass[pre,twocolumn,english,superscriptaddress,floatfix,longbibliography,nofootinbib]{revtex4-2}

\usepackage[normalem]{ulem}
\usepackage{graphicx}
\usepackage{dcolumn}
\usepackage{bm}
\usepackage{xcolor}
\usepackage[dvipsnames]{xcolor}
\usepackage{amsmath}
\usepackage{amssymb}

\usepackage{amssymb}
\usepackage{lineno}
\usepackage{physics, mathtools}

\usepackage{xcolor}

\usepackage{hyperref}
\hypersetup{colorlinks=true, citecolor=blue, linkcolor=black, urlcolor=blue}

\begin{document}

\date{\today}


\title{Adaptive thresholding for scalable measurement-based qubit reset}

 \author{Qian Cao}
\affiliation{Department of Electrical Engineering and Computer Science, University of California Berkeley, Berkeley, CA, USA, 94720.}
 \affiliation{Department of Physics, Washington University, St. Louis, Missouri 63130, USA}
\author{Unnati Akhouri}
 \affiliation{Quantum Machines, Boston, USA}
\author{Samuel Vizvary}
 \affiliation{Quantum Machines, Los Angeles, USA}
\author{Nissim Ofek}
 \affiliation{Quantum Machines, Tel Aviv, Israel}
\author{Wei Dai}
\email{wei.dai@quantum-machines.co}
\affiliation{Quantum Machines, Boston, USA}
\author{Kater W. Murch}%
\email{katermurch@berkeley.edu}
\affiliation{Department of Electrical Engineering and Computer Science, University of California Berkeley, Berkeley, CA, USA, 94720.}
\affiliation{Department of Physics, University of California Berkeley, Berkeley, CA, USA, 94720.}
 \affiliation{Department of Physics, Washington University, St. Louis, Missouri 63130, USA}

\begin{abstract}
Fast, high-fidelity qubit initialization is a key primitive for scalable quantum information processing, but conventional reset protocols either require long relaxation times or discard information contained in continuous measurement records. We demonstrate an adaptive measurement-based reset protocol for superconducting qubits that uses Bayesian inference to update the qubit-state estimate after each readout and dynamically adjust the feedback threshold. Unlike fixed-threshold or repeat-until-success (RUS) protocols, the method utilizes the full analog measurement history, enabling the reset decision to become progressively more conservative as confidence in ground-state preparation increases. Implemented with real-time FPGA-based feedback on a dispersively readout transmon qubit, the protocol achieves a ground-state initialization fidelity of $99.44\pm0.04\%$ after a small number of reset rounds. We further compare adaptive reset with RUS strategies in multi-qubit experiments and show that adaptive thresholding provides deterministic reset duration while maintaining a five-qubit simultaneous-reset fidelity of $98.78 \pm 0.19\%$. These results establish Bayesian adaptive thresholding as a practical and scalable route to fast qubit reset in superconducting quantum processors. 
\end{abstract} 

\maketitle

\section{Introduction}
Rapid and high-fidelity initialization of qubits into a known fiducial state is a foundational primitive that underlies quantum algorithms~\cite{DiVincenzo2000,Nielsen_qcqi} and quantum error correction~\cite{Fowler2012,Schindler2011,Reed2012}. 
As quantum processors scale, reset increasingly becomes an active resource rather than a passive operation: long initialization times reduce computational throughput, while imperfect reset introduces residual excited-state population that propagates errors into subsequent operations~\cite{McEwen2021,Jin2015,Egger2018}. 
In superconducting quantum processors, initialization is often performed by waiting several energy relaxation times for the system to return to thermal equilibrium~\cite{Rigetti2012,Place2021}. 
While conceptually simple, passive reset cannot be used for mid-circuit reset operations. Moreover, the residual thermal population sets a fundamental limitation on the achievable initialization fidelity. 
These challenges have motivated the development of reset protocols that actively prepare the qubit in its ground state on demand \cite{Salath2018,
Magnard2018,Zhou2021}.

A broad range of active reset techniques have been developed, which we categorize based on whether they rely on measurements to remove entropy. 
Non-measurement-based reset schemes generally involve enhancing energy relaxation rates, by resonant Purcell decay~\cite{Reed2010,tuorilaEfficientProtocolQubit2017,McEwen2021}, dynamical coupling to lossy modes~\cite{Geerlings2013,Magnard2018,Zhou2021,Yoshioka2023}, or reservoir engineering~\cite{Murch2012,Aamir2025}. 
The reset fidelity of these schemes is eventually limited by the bath temperature. 
Measurement-based reset schemes, which rely on high-fidelity, quantum non-demolition (QND) readout, can achieve heralded qubit initialization~\cite{Rist2012, Johnson2012}, or on-demand reset with real-time feedback~\cite{Salath2018,Tholn2022}. 

Measurement-based reset protocols typically convert each readout outcome into a binary assignment of the qubit state using a fixed  threshold. Their reset fidelity is therefore limited by single-shot readout errors. Repeating the measurement and feedback cycle can improve the reset fidelity by acquiring additional information about the qubit state, as in repeat-until-success (RUS) protocols. However, the number of required rounds is then stochastic, resulting in a non-deterministic reset duration. This limits the scalability of simultaneous multi-qubit reset, since the total reset time is determined by the slowest qubit and therefore tends to increase with the number of reset qubits. More generally, conventional measurement-based reset protocols do not fully utilize the information contained in the measurement records. By reducing each measurement outcome to a binary decision, these protocols discard both the confidence of each outcome and information accumulated over multiple rounds.

In this work, we demonstrate an adaptive measurement-based reset protocol that retains this information by continuously updating the reset decision using the full analog measurement history. Rather than applying fixed decision boundaries, our approach uses Bayesian inference to estimate the probability that the qubit occupies $\ket{g}$ after each measurement and dynamically adjusts the feedback threshold accordingly. We experimentally demonstrate that this approach overcomes two key limitations of prior measurement-based reset techniques: First, our reset protocol is fast and time-deterministic. This enables simultaneous multi-qubit reset.  Second, the protocol is robust to low readout signal-to-noise ratio (SNR). These results enable improved initialization fidelity for a given reset duration and illustrate how adaptive processing of continuous quantum measurement records can enhance feedback-based quantum control.

The paper is organized as follows: In Section II, we analyze the conventional measurement-based reset scheme and a RUS scheme for pedagogical purpose. In Section III, we explain the adaptive thresholding reset scheme, and experimentally demonstrate its performance. 
In Section IV, we discuss the scalability of our scheme, and experimentally compare it with the RUS scheme for resetting multiple qubits.

\section{Measurement-based qubit reset}

\begin{figure}
\includegraphics[width=.8\columnwidth]{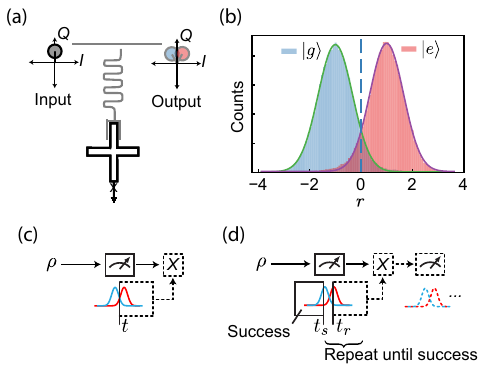}
\caption{
\textbf{
Measurement-based active reset using dispersive qubit readout. }
(a) Standard dispersive measurement of a transmon qubit coupled to a microwave resonator. The qubit-state-dependent frequency shift of the resonator maps the qubit state onto the phase of a transmitted coherent probe field. 
(b) Histograms of the integrated measurement outcome $I$ obtained following preparation of $\ket{g}$ and $\ket{e}$. The distributions are approximately Gaussian with means $\mu_g$ and $\mu_e$ and variances $\sigma_g^2$ and $\sigma_e^2$, and the outcomes are scaled such that $\mu_g=-1$ and $\mu_e=+1$. The separation and overlap determine the measurement strength and readout fidelity. 
(c) Conventional measurement-based active reset. For an arbitrary initial state $\rho$, a projective measurement is performed and a conditional $X_\pi$ pulse is applied when the measurement result exceeds a threshold, ideally preparing the qubit in $\ket{g}$. The achievable reset fidelity is limited by readout errors. 
(d) Multi-round reset with independent measurement and reset thresholds. One threshold is used to herald successful preparation of $\ket{g}$, while a second threshold determines whether a conditional $X_\pi$ operation is applied. The protocol is repeated until the success (RUS) condition is reached, enabling initialization fidelities beyond the single-shot readout fidelity at the expense of a variable number of measurement rounds.
}
\label{fig:fig1}
\end{figure}

The information used by the adaptive protocol is naturally available in dispersive qubit readout. In this setting, the qubit state is mapped onto an analog pointer signal, which can be calibrated, processed in real time, and converted into conditional control operations. 
Although we implement the protocol with a superconducting transmon qubit coupled to a microwave resonator, the same information structure appears more broadly in measurement platforms based on dispersive coupling to microwave or optical pointer states. The basic readout architecture is illustrated in Fig.~\ref{fig:fig1}(a): a coherent probe interrogates a resonator whose response depends on the qubit state, producing a continuous-valued measurement outcome rather than an intrinsically binary result.

In the dispersive regime, the qubit state shifts the resonance frequency of the readout resonator through the interaction Hamiltonian
\begin{equation}
H_{\rm disp}/\hbar=-\chi a^\dagger a\sigma_z ,
\end{equation}
where $\chi$ is the dispersive coupling strength, $a^\dagger a$ is the resonator photon number operator, and $\sigma_z=\ket{g}\bra{g}-\ket{e}\bra{e}$ with $\ket{g}$ ($\ket{e}$) denoting the ground (first excited) state of the transmon. Consequently, a coherent microwave probe interacting with the resonator acquires a qubit-state-dependent phase response, allowing information about the qubit state to be encoded in the transmitted or reflected measurement signal. After amplification, demodulation, and integration of the transmitted signal, each measurement produces a continuous-valued outcome $r\in\mathbb{R}$ along the principal axis denoted as the $I$ quadrature.

To characterize the measurement, we prepare the qubit in either $\ket{g}$ or $\ket{e}$ and record the corresponding distributions of integrated readout outcomes, shown  in Fig.~\ref{fig:fig1}(b). We scale the readout such that the mean values of the two distributions take values $\mu_g=-1$ and $\mu_e=+1$. 
The measurement outcomes are well approximated by Gaussian distributions,
\begin{equation}
P(r|g)
=
\frac{1}{\sqrt{2\pi\sigma_g^2}}
\exp\!\left[
-\frac{(r-\mu_g)^2}{2\sigma_g^2}
\right],
\end{equation}

\begin{equation}
P(r|e)
=
\frac{1}{\sqrt{2\pi\sigma_e^2}}
\exp\!\left[
-\frac{(r-\mu_e)^2}{2\sigma_e^2}
\right].
\end{equation}
The signal-to-noise ratio $\mathrm{SNR}\equiv|\mu_e-\mu_g|/\sqrt{\sigma_g^2+\sigma_e^2}$ determines
the degree of overlap between the two distributions, which further determines the single-shot readout assignment fidelity. 
As an example, the distributions shown in Fig.~\ref{fig:fig1}(b) yield an SNR of 2.11, resulting in state assignment fidelity of approximately $92\%$. 
This moderate-fidelity operating point is chosen to make the multi-round evolution of the Bayesian prior and adaptive threshold clearly visible.  At high SNR, the overlap between the two distributions becomes small such that the adaptive dynamics become less pronounced. The reduced readout contrast used here should therefore be viewed as an illustrative operating point rather than a technical limitation. State-of-the-art dispersive readout of superconducting transmon qubits can achieve single-shot assignment fidelities above $99\%$ within sub-microsecond measurement times ~\cite{Walter2017,Chen2023,https://doi.org/10.48550/arxiv.2601.04975}.

The measurement outcome can be used directly to initialize the qubit through feedback. 
In the conventional protocol shown in Fig.~\ref{fig:fig1}(c): if the measurement outcome $r$ exceeds a fixed threshold $t$, a conditional $X_\pi$ pulse is applied to invert the qubit population and prepare $\ket{g}$. This procedure converts measurement information into entropy removal and can substantially reduce the reset time compared to passive relaxation. 
However, because the feedback decision is based on a single noisy observation, the achievable initialization fidelity remains fundamentally constrained by the readout assignment errors.

To overcome this limitation, a RUS protocol is designed to repeat measurement and feedback until a sufficiently informative outcome heralds successful preparation of $\ket{g}$, as illustrated in Fig.~\ref{fig:fig1}(d). Two thresholds partition the readout space $r$ into three distinct regions. The outer regions correspond to a high probability of the qubit being in the ground state $|g\rangle$ or excited state $|e\rangle$, respectively. Measurements that fall within the intermediate region are considered ambiguous, and additional measurements are performed to resolve the state assignment. Accordingly, separate thresholds can be used to determine whether a conditional $X_{\pi}$ pulse is applied and whether the reset sequence terminates. This approach can achieve initialization fidelities exceeding the nominal single-shot assignment fidelity. The tradeoff is that the number of measurement rounds becomes stochastic, increasing the average reset duration. Particularly, the expected number of rounds scales with the number of qubits to reset, making the repeat-until-success protocol not scalable in resetting a large quantum processor.
A detailed analysis of the repeat-until-success protocol and its dependence on the two thresholds is given in App.~\ref{App:RUS}. 
While the RUS protocol harnesses the confidence encoded within each individual measurement outcome, it discards the information acquired in previous rounds and therefore does not leverage the cumulative evidence gathered over the multiple measurement rounds. In the following, we instead retain and accumulate information across the full measurement history and adapt the reset decision dynamically.

\begin{figure}
\includegraphics[width=\columnwidth]{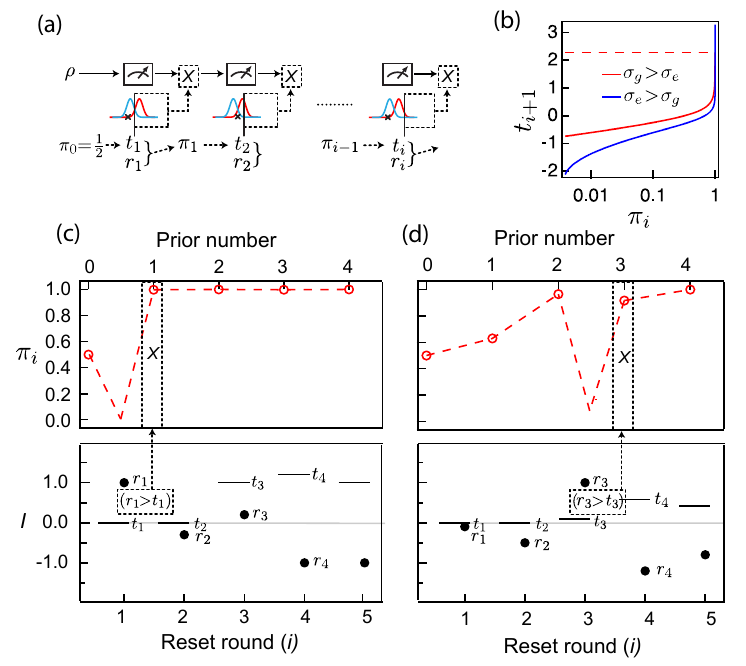}
\caption{
\textbf{Adaptive-threshold reset protocol based on Bayesian updating of the qubit state estimate.} 
(a) Sequence of repeated measurements and conditional feedback operations. At round $i$, the prior probability $\pi_{i-1}=P(g)$ is used to determine the measurement threshold $t_{i}$. Following acquisition of measurement outcome $r_i$ (denoted by $\times$ symbols), the posterior probability $\pi_{i}$ is updated and used to compute the threshold for the subsequent round. Conditional reset operations are applied according to the updated state estimate. 
(b) Reset threshold $t_{i+1}$ as a function of the prior probability $\pi_i$ for the cases $\sigma_g > \sigma_e$ and $\sigma_e > \sigma_g$, illustrating how the adaptive decision boundary is updated during the reset protocol. 
(c,d) Representative trajectories of the adaptive reset process. Lower panels show the measurement outcomes $r_i$ and corresponding thresholds $t_i$ at each round, while upper panels show the evolution of the posterior ground-state probability $\pi_i$. When the measurement result satisfies the reset criterion ($r_i>t_i$), a conditional $X_\pi$ pulse is applied and the state estimate is updated accordingly. 
}
\label{fig:fig2}
\end{figure}

\section{Adaptive thresholding for reset}
\subsection{Bayesian adaptive thresholding}
The adaptive reset protocol is illustrated in Fig.~\ref{fig:fig2}(a). Rather than maintaining a binary estimate of the qubit state, after each measurement we track a continuously updated prior probability, $\pi_i$ that quantifies our confidence that the qubit occupies $\ket{g}$ after round $i$.

Beginning with an initial prior $\pi_0$ (typically assumed to be $1/2$), each measurement outcome $r_i$ ($i = 1, 2,3...$) updates the state estimate according to Bayes' rule,
\begin{equation}
\pi_i
=
\frac{
\pi_{i-1}P(r_i|g)
}{
\pi_{i-1}P(r_i|g)
+
(1-\pi_{i-1})P(r_i|e)
},
\end{equation}
which preserves all information contained in the analog measurement record.

The feedback threshold is then chosen adaptively. We define the threshold $t_{i+1}$ as the measurement value for which the prior probabilities of the two states are equal,
\begin{equation}\label{eq:threshold}
P(t_{i+1}|g)\,\pi_{i}
=
P(t_{i+1}|e)\,(1-\pi_{i}).
\end{equation}

Solving this condition yields a threshold $t_{i+1}$ that depends explicitly on the prior confidence $\pi_i$. Figure~\ref{fig:fig2}(b) shows the resulting threshold as a function of prior probability for typical readout distributions as in Fig.~\ref{fig:fig1}(b). $t_{i+1}$ is an increasing function of $\pi_i$; as confidence that the qubit is already in $\ket{g}$ increases, progressively larger measurement outcomes are required to compensate the Bayesian prior and trigger the application of a $X_\pi$ rotation. When $\sigma_g > \sigma_e$, both the prior probability $\pi_i$ and the corresponding threshold $t_{i+1}$ have finite upper bounds. In contrast, when $\sigma_g < \sigma_e$, the threshold $t_{i+1}$ diverges as $\pi_i$ approaches 1. The Bayesian update and threshold calculation for general Gaussian readout distributions with unequal variances are derived in App.~\ref{App.theory calculation}.

Figure~\ref{fig:fig2}(c,d) display example trajectories of the reset procedure.  In Fig.~\ref{fig:fig2}(c), the first measurement outcome exceeds the initial threshold and immediately triggers an $X_\pi$ operation. Subsequent measurements increase confidence that the qubit occupies $\ket{g}$, causing the threshold to shift toward more conservative reset decisions. Figure~\ref{fig:fig2}(d) illustrates a representative trajectory in which the first two measurement outcomes lie close to the decision boundary and are therefore not strongly indicative of either state. However, these outcomes update the prior based on Bayesian inference and, consequently, cause the decision threshold for triggering the $X_\pi$ pulse to evolve. In the third round, the observed measurement outcome falls on the excited-state side of the updated threshold, where sufficient confidence is obtained to trigger the $X_\pi$ pulse. 

These examples illustrate the central idea of adaptive reset: the protocol uses the full analog measurement history to continuously update both the estimated qubit state and the decision rule itself. The corresponding reduction in uncertainty of the Bayesian state estimate and its connection to information gained from successive measurements are discussed in App.~\ref{app:entropy}.

\begin{figure}
\includegraphics[width=1\columnwidth]{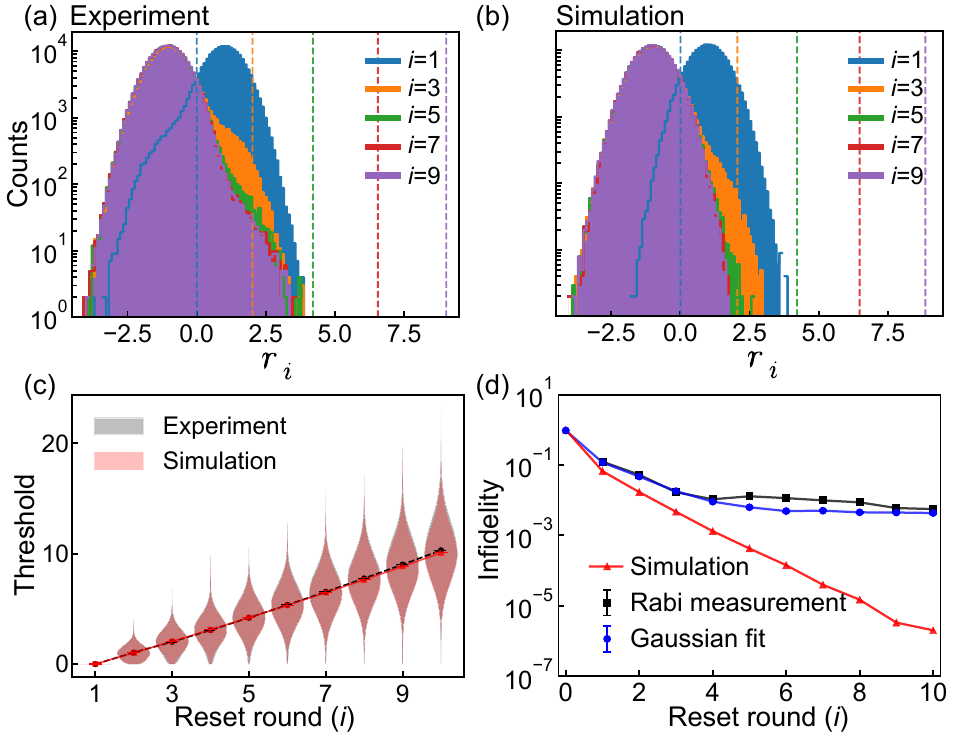}
\caption{
\textbf{Performance of adaptive threshold reset on a single superconducting qubit.} 
(a) Distributions of measurement outcomes $r_i$ following successive rounds of adaptive reset, beginning from $\ket{e}$. After the first round, conditional feedback suppresses the population in $\ket{e}$, producing progressively stronger weighting of the measurement distribution toward outcomes associated with $\ket{g}$. As the posterior probability $\pi_i$ approaches unity, the reset thresholds (vertical dashed lines) shift toward more conservative values, reflecting increasing confidence that the qubit occupies $\ket{g}$. 
(b) Distributions of the measurement outcomes $r_i$ obtained from Monte Carlo simulations of the adaptive-reset protocol. The simulation assumes quantum-nondemolition (QND) measurements and applies the same Bayesian threshold-update and conditional-feedback procedure as in the experiment. The simulated distributions reproduce the progressive accumulation of population in $\ket{g}$ over successive reset rounds.
(c) Validation of the reset-threshold distribution. The violin plots show the distributions of the reset threshold at each round. Black and red markers indicate the experimental and Monte Carlo simulation results, respectively. The black dashed line and red solid line connect the corresponding mean threshold values. The close agreement between the experimental and simulated distributions validates the threshold-update protocol.
(d) $\ket{g}$-state initialization infidelity versus reset round $i$ ($i = 0$ stands for the initially prepared state). Blue markers show the infidelity extracted from the two-Gaussian fit, black markers show the infidelity measured using the $\ket{e}\leftrightarrow\ket{f}$ Rabi method \cite{Geerlings2013}, and red markers show the infidelity obtained from Monte Carlo simulations. The measured initialization infidelity decreases rapidly during the first few reset rounds and subsequently approaches a floor of approximately $6\times10^{-3}$, corresponding to a maximum initialization fidelity of $99.44\pm0.04\%$.} 
\label{fig:fig3}
\end{figure}

\begin{figure}
\includegraphics[width=\columnwidth]{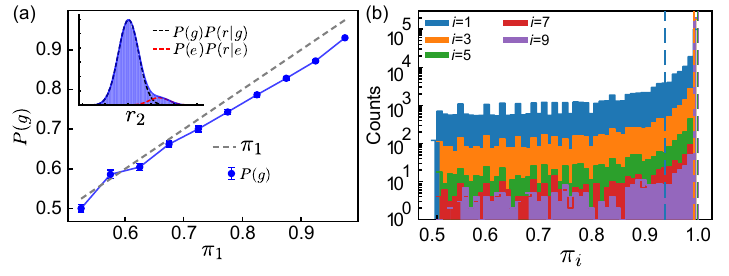}
\caption{
\textbf{Prior probability distribution.} (a) Comparison between the estimated prior $\pi_i$ of the first round and $P(g)$ inferred from fitting following measurement distribution (inset). The measurement results $r$ are binned according to $\pi_i$ for fitting (blue points). The dashed line indicates unit slope. (b) The distribution of $\pi_i$ after each reset cycle.
}
\label{fig:fig4}
\end{figure}
\subsection{Experimental demonstration}
To experimentally demonstrate adaptive threshold reset, we implement the protocol on a superconducting transmon qubit with $\ket{g}\leftrightarrow\ket{e}$ transition frequency $\omega_\mathrm{q}/2\pi = 4.076$ GHz dispersively coupled to a microwave readout resonator with resonance frequency $\omega_\mathrm{r}/2\pi = 7.644$ GHz. Readout is performed using coherent microwave pulses applied to the resonator and detecting the transmitted response. The outgoing measurement signals are amplified using a traveling-wave parametric amplifier (TWPA) operated near the quantum limit~\cite{https://doi.org/10.48550/arxiv.2406.19752}, followed by additional cryogenic and room-temperature amplification stages. The amplified signals are heterodyne demodulated and integrated to produce the continuous-valued measurement record $r$ introduced in Fig.~\ref{fig:fig1}. Additional details of the device and parameters for all five qubits are provided in App.~\ref{app:device}.

Real-time feedback is implemented using a Field-Programmable Gate Array (FPGA)-based OPX$+$ controller, which digitizes the measurement outcomes, evaluates the Bayesian update rule, computes the adaptive threshold, and conditionally applies reset pulses with low latency~\cite{Berritta2026,https://doi.org/10.48550/arxiv.2602.11912,Dassonneville2026}. Following each measurement round $i$, the controller updates the posterior probability $\pi_i=P(g)$ from the measured value $r_i$, evaluates the threshold condition derived in the previous section, and determines whether a conditional $X_\pi$ pulse should be applied before proceeding to the next round. This architecture allows the adaptive reset protocol to be executed entirely in hardware without offline processing.  In our implementation with the OPX$+$ controller, the total latency required to update the prior probability before applying the conditional $X_{\pi}$ pulse is 428~$\mathrm{ns}$ (App.~\ref{app:timing} contains a detailed timing diagram). This latency is dominated by the signal time-of-flight, 288~$\mathrm{ns}$, while the FPGA processing time for updating the prior probability is 140~$\mathrm{ns}$. After the conditional $X_{\pi}$ pulse, the FPGA requires an additional 24~$\mathrm{ns}$ to compute the threshold for the next measurement round. In our experiment, a depletion time longer than $1 ~\mu\mathrm{s}$ is applied after each measurement to allow photons in the readout resonator to decay. This depletion time is much longer than the FPGA processing latency and therefore effectively covers the time required for the prior update.

Before performing the adaptive reset, we first extract two Gaussian distributions from the measured ground- and excited-state readout histograms. For the first measurement round, since the initial state is unknown, we set the prior probability to $\pi_0 = 0.5$ and calculate the initial threshold in advance using Eq.~\ref{eq:threshold}. After each measurement round $i$, the measured result $r_i$ is used to update the posterior probability. If $r_i > t_i$, corresponding to a posterior probability $\pi_i > 0.5$, an $X_\pi$ pulse is applied to reset the qubit, and the posterior probability is correspondingly updated to $1-\pi_i$. Using this updated posterior probability, we then solve for $t_{i+1}$, the threshold for the next measurement round, that satisfies Eq.~\ref{eq:threshold}. When the ground-state Gaussian has a larger standard deviation than the excited-state Gaussian, $\sigma_g>\sigma_e$, the prior probability has an upper bound given by
\begin{equation}
\pi_{\max}
=
\frac{1}{
1+\frac{\sigma_e}{\sigma_g}
\exp\left[
-\frac{(\mu_g-\mu_e)^2}{2(\sigma_g^2-\sigma_e^2)}
\right]
}.
\label{eq:pi_max}
\end{equation}
Once this maximum value is reached, we keep the threshold fixed in the following rounds instead of continuing to update it. Beyond this bound, the $\ket{g}$ distribution covers the $\ket{e}$ distribution. Further measurements can not provide sufficient evidence to favor $\ket{e}$, so a new decision threshold can no longer be defined.

Figure~\ref{fig:fig3}(a) shows the evolution of the measured readout distributions over successive rounds of adaptive reset beginning from the state of $\ket{e}$.  The initial distribution therefore exhibits weight peaked at the excited state distribution. The residual population near the $\ket{g}$ distribution arises from $T_1$ decay during the measurement. After the first round of adaptive reset, the weight associated with $\ket{e}$ is visibly reduced, indicating successful transfer of excited-state population toward the ground state. With each subsequent round, the distribution becomes increasingly concentrated near outcomes corresponding to $\ket{g}$.  The vertical dashed lines indicate the average adaptive threshold applied at each round and reveal a second effect beyond population transfer. As information accumulates and the posterior probability $\pi_i$ approaches unity, the threshold shifts toward increasingly conservative values. Early in the reset sequence, thresholds remain permissive in order to aggressively remove excited-state population. At later rounds, once the qubit is believed with high confidence to occupy $\ket{g}$, reset operations are applied only when supported by increasingly strong measurement evidence.

To verify that the adaptive-reset protocol operates as intended, we perform Monte Carlo simulations using the same number of trajectories as experimental shots. In each trajectory, the measurement outcomes are sampled from the calibrated $\ket{g}$- and $\ket{e}$-state readout distributions under the assumption of quantum-nondemolition (QND) measurement. The same Bayesian threshold-update rule and conditional-feedback procedure used in the experiment are then applied. The simulated measurement distributions reproduce the progressive redistribution of the outcomes toward the $\ket{g}$-state distribution over successive reset rounds [Fig.~\ref{fig:fig3}(b)]. Moreover, the simulated threshold distributions and their mean values closely agree with the experimental results [Fig.~\ref{fig:fig3}(c)], confirming the correct implementation of the Bayesian threshold-update protocol. The statistical distribution of the adaptive threshold and its evolution over successive reset rounds are analyzed in App.~\ref{app:threshold-dist}.

To quantify the performance of the adaptive reset protocol, we directly measure the resulting initialization infidelity as a function of the number of reset rounds, shown in Fig.~\ref{fig:fig3}(d). Because the reset fidelity exceeds the assignment fidelity of a single readout, direct characterization from the $\ket{g}$ and $\ket{e}$ measurement distributions becomes unreliable. Instead, we employ the population measurement technique introduced in Ref.~\cite{Geerlings2013}, which infers residual excited-state occupation by comparing the amplitudes of Rabi oscillations on the $\ket{e}\leftrightarrow\ket{f}$ transition. This method provides sensitivity to excited-state populations below the single-shot readout error floor. We also compare the initialization infidelity extracted from a two-Gaussian fit. Specifically, we fit the measurement outcomes $r_i$ to a weighted mixture of two Gaussian distributions, $P(r|g)$ and $P(r|e)$. The infidelities obtained from the two-Gaussian fit are consistent with those measured using the $\ket{e}\leftrightarrow\ket{f}$ Rabi method.  The measured initialization fidelity improves rapidly over the first several rounds of adaptive reset before approaching saturation after approximately six rounds. Early rounds primarily remove residual excited-state population and therefore yield substantial gains in fidelity. In later rounds, the qubit is already inferred with high confidence to occupy $\ket{g}$, and the adaptive threshold becomes increasingly conservative, suppressing unnecessary feedback operations and leading to diminishing returns from additional reset cycles. The idealized Monte Carlo simulation predicts a continued reduction in initialization infidelity, whereas the experimental infidelity reaches a floor of approximately $6\times10^{-3}$. The observed saturation in the experimental data therefore results from a combination of experimental details not captured by the model,  including measurement-induced state transitions (MISTs), population in the higher levels of the transmon, and relaxation and thermal excitation during the finite-duration measurement.

\begin{figure*}
\includegraphics[width=2\columnwidth]{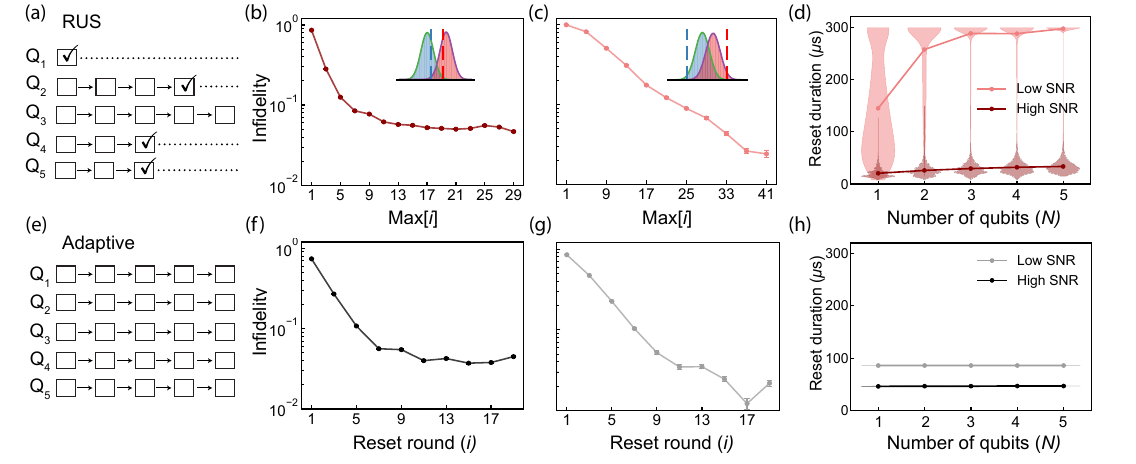}
\caption{
\textbf{Scaling of reset protocols for multi-qubit state preparation.} 
(a)--(d) Repeat-until-success reset applied to a five qubit array. (a) Each qubit undergoes measurement and conditional reset operations until a successful initialization event is heralded (check marks). (b) Reset infidelity as a function of the maximum allowed number of reset cycles, $\mathrm{Max}[N]$, in the high-SNR regime. The inset illustrates the two readout thresholds, chosen such that the $\ket{g}$ and $\ket{e}$ state outcomes each have a state purity of $99\%$. (c) Reset infidelity as a function of $\mathrm{Max}[N]$ in the low-SNR regime, obtained using half the readout amplitude. 
(d) Mean reset duration and spread in reset times for increasing qubit number of RUS reset protocols. The violin plots indicate the distribution of reset durations, while the markers and connecting lines indicate their means. Dark red and light red denote the high- and low-SNR regimes, respectively. The timing distributions broaden and the mean reset duration increases with qubit number, particularly in the low-SNR regime.
(e)--(h) Adaptive threshold reset applied to multiple qubits. (e) The number of measurement rounds $N$ is fixed in advance, resulting in deterministic protocol duration. (f) Reset infidelity as a function of the number of reset rounds in the high-SNR regime, showing saturation of the reset performance after repeated adaptive updates. (g) Reset infidelity versus $N$ in the low-SNR regime.  
(h) Mean reset duration and spread in reset times for increasing qubit number of adaptive  reset protocols. Black and gray points denote the high- and low-SNR regimes, respectively. Because the adaptive protocol uses a fixed number of reset rounds, its total duration remains constant as the number of qubits increases.
}
\label{fig:fig5}
\end{figure*}

A key assumption underlying the adaptive protocol is that the posterior probability $\pi_i$ provides a faithful estimate of the true qubit state. To verify this, we compare the Bayesian estimate against an independent ground truth obtained by partitioning the measurement record into bins according to $\pi_i$ and fitting the conditional measurement distributions within each bin. Figure~\ref{fig:fig4}(a) shows the inferred $P(g)$ extracted from Gaussian fits to the binned distributions plotted against the corresponding Bayesian estimate $\pi_i$ in the first reset round. The results lie close to the line of unit slope across the full range of prior values, confirming that the Bayesian update rule accurately tracks the true state occupation probability throughout the protocol. Small deviations near $\pi_i \approx 1$ are consistent with the finite residual excited-state population remaining after many reset rounds.

Figure~\ref{fig:fig4}(b) shows the distribution of $\pi_i$ values across reset rounds. At the start of the protocol the prior is broadly distributed, reflecting genuine uncertainty in the qubit state. With successive rounds, the distribution accumulates sharply near $\pi_i \approx 1$, indicating that the protocol rapidly drives the ensemble toward high-confidence ground-state preparation. Together, these results confirm that the Bayesian state estimate serves as a reliable proxy for the true qubit state and provides a sound basis for the adaptive threshold decisions.

\section{Multi-qubit reset}
A practical consideration for near-term quantum processors is that qubit initialization must scale to arrays of many qubits simultaneously. While the adaptive protocol operates independently on each qubit, its fixed-duration structure offers a distinct practical advantage over RUS approaches when applied in parallel. To enable a direct comparison among the protocols, we set the duration of each reset round to be $7~\mu$s for all qubits, using the same readout duration $2~\mu$s and depletion time $5~\mu$s. Figure~\ref{fig:fig5} compares the performance of RUS and adaptive threshold reset as a function of the number of qubits reset simultaneously.

Figure~\ref{fig:fig5}(a) illustrates the RUS strategy applied to a five-qubit array. In this scheme, each qubit undergoes repeated measurement and conditional reset until a successful initialization event is heralded. Because the number of RUS rounds required for each qubit is stochastic, each qubit has a nondeterministic reset duration. In resetting a multi-qubit quantum processor, success must be achieved simultaneously across all qubits. The total reset duration is therefore set by the slowest qubit in each run. The two thresholds are chosen such that both the $\ket{g}$ and $\ket{e}$ state purities reach 0.99 [see Appendix~\ref{App:RUS}]. As shown in Fig~\ref{fig:fig5}(b) and (c), we compare the RUS reset protocol in the high- and low-signal-to-noise ratio regimes. In the high-SNR regime, the single-qubit readout SNRs for $Q_1–Q_5$ are $2.22$, $2.11$, $2.09$, $2.65$, and $2.14$, respectively. The low-SNR regime is obtained by reducing the readout amplitude by a factor of two, yielding corresponding SNRs of $1.12$, $1.05$, $1.03$, $1.35$, and $1.05$. The low-SNR regime requires significantly more reset cycles to achieve comparable reset fidelity, because the reduced measurement contrast lowers the efficiency of the threshold-based decision. Figure~\ref{fig:fig5}(d) quantifies the distribution of reset durations as the number of qubits increases under the RUS protocol. We set the maximum number of allowed reset cycles to $20$ in the high-SNR regime and $40$ in the low-SNR regime, such that the average reset fidelity (measured by gaussian fit) of each qubit remains above $98\%$. In both regimes, the mean reset duration increases with qubit number. This increase is substantially more pronounced in the low-SNR regime, where the reset-duration distribution progressively shifts toward the maximum allowed duration.

Figure~\ref{fig:fig5}(e) shows the adaptive threshold protocol applied to the same multi-qubit setting with a fixed number of reset rounds $N$ per qubit. Because the protocol duration is deterministic, all qubits complete reset simultaneously regardless of the individual measurement trajectories, eliminating the timing overhead that accumulates in RUS schemes. The five-qubit simultaneous reset fidelity reaches a maximum of $96.27\pm0.08\%$ in the high-SNR regime and $98.78\pm0.19\%$ in the low-SNR regime as the number of reset rounds $N$ increases, as shown in Fig.~\ref{fig:fig5}(f) and (g). The number of rounds required to reach this fidelity is similar in the two cases.
Figure~\ref{fig:fig5}(h) shows the reset duration under the adaptive reset protocol. We use $6$ adaptive reset rounds in the high-SNR regime and $11$ rounds in the low-SNR regime, such that the average reset fidelity of each qubit exceeds approximately $98\%$.

A direct comparison of the two protocols reveals a trade-off between measurement SNR and the maximum achievable reset fidelity. For both RUS and adaptive reset, the low-SNR regime reaches a slightly higher maximum fidelity than the high-SNR regime, despite requiring more reset rounds. This improvement arises because the reduced readout amplitude suppresses MISTs \cite{Khezri2023,Sank2016,Nesterov2024,Dai2026}, which impose a fidelity floor at higher measurement power. The adaptive protocol is particularly advantageous in the low-SNR regime: compared with RUS, it achieves a substantially shorter mean reset duration while eliminating the broad, qubit-number-dependent timing distribution associated with the random number of reset cycles. These results show that adaptive-threshold reset provides a scalable initialization primitive for multi-qubit processors where both high fidelity and timing determinism are required.

\section{Conclusion and outlook}
We have demonstrated an adaptive measurement-based reset protocol that uses Bayesian inference to process the full analog readout history of a superconducting qubit. Unlike fixed-threshold reset or RUS protocols, the adaptive protocol continuously updates the estimated ground-state probability and adjusts the feedback threshold after each measurement round. As confidence in ground-state preparation increases, the protocol naturally becomes more conservative, suppressing unnecessary reset operations while preserving high initialization fidelity. In single-qubit experiments, this approach achieves high-fidelity initialization after a small number of measurement rounds. In multi-qubit reset, the fixed-round structure provides deterministic timing, avoiding the stochastic duration and increasing timing spread associated with RUS strategies. These results establish adaptive thresholding as a practical method for fast qubit initialization and, more generally, demonstrate the value of treating analog measurement records as likelihood-bearing resources for real-time quantum control. A natural extension is to generalize the protocol to multilevel systems by retaining the full two-dimensional IQ measurement record and performing Bayesian inference over multiple states, enabling adaptive reset of qutrits or higher-dimensional systems.

The same principle can be extended beyond reset to other settings in which noisy measurement records are converted into control decisions. Related ideas of combining repeated measurement information with Bayesian feedback have also been explored beyond superconducting circuits~\cite{Kobayashi2023}. In quantum error correction, for example, syndrome measurements are typically reduced to binary outcomes, even though the underlying analog readout contains confidence information that can improve decoding performance. Recent work on soft-information decoding has shown that retaining analog syndrome information can improve error-correction performance in both theoretical models and superconducting-qubit experiments~\cite{2107.13589,Ali2024,Hanisch2026}. Adaptive processing of measurement records may therefore provide a useful hardware-level primitive for real-time decoders, Pauli-frame updates, and feedback-enabled error correction, where low-latency classical processing is increasingly central to scalable architectures~\cite{2024,Battistel2023}. More broadly, Bayesian decision rules could be incorporated into measurement-based feedback and feed-forward protocols, including state stabilization, trajectory-dependent dynamics, and measurement-conditioned state preparation~\cite{Vijay2012,Zhang2017}. Similar ideas may also be useful for the calibration and operation of large-scale quantum processors, where controllers must decide when sufficient information has been acquired, whether a parameter update is warranted, and how to branch efficiently through many calibration tasks~\cite{2104.10866,2205.12929,Alexeev2025}. In this broader view, adaptive reset is one example of a general strategy for embedding real-time statistical inference into quantum control hardware.

\begin{acknowledgments}
This work received support from the National Science Foundation award No.~PHY-2408932 and ONR Grant No.~N000142512160. The qubit device was fabricated and provided by the Superconducting Qubits at Lincoln Laboratory (SQUILL) Foundry at MIT Lincoln Laboratory, with funding from the Laboratory for Physical Sciences (LPS) Qubit Collaboratory.
\end{acknowledgments}

\section*{Author contributions}
Kater Murch and Qian Cao designed the experiments. 
Qian Cao Performed the experiments, data curation and analysis. 
Wei Dai and Nissim Ofek contributed the initial idea of adaptive thresholding for reset. 
Unnati Akhouri formalized the theoretical analysis underlying the protocols, developed and verified the theoretical and empirical results. 
Nissim Ofek developed the controller architecture that enabled adaptive thresholding. 
Sam Vizvary and Wei Dai assisted in implementing the moving threshold on the OPX controller. 
Kater Murch and Wei Dai oversaw the project. 
All authors contributed to the writing and presentation of the results. 

\section*{Data Availability}
The data and code that support the findings of this article are openly available at \url{https://doi.org/10.5281/zenodo.22719098}.
\bibliography{refs_cleaned}

\section*{Appendix}

\appendix
\section{Repeat Until Success Protocol}
\label{App:RUS}

The repeat-until-success (RUS) protocol for active qubit reset is designed to mitigate residual errors caused by the overlap of two Gaussian distributions. In standard active reset, a single threshold is used to distinguish qubit states, but the overlap region between the Gaussians leads to ambiguous outcomes and misassigned states, lowering fidelity. As illustrated in Fig.~\ref{fig:fig1}, there is a nonzero probability for an excited-state outcome to be misidentified as ground state and vice versa, particularly for measurement outcomes near the threshold. This limitation, set by the overlap area of the two Gaussians, bounds the achievable single-shot fidelity.

The RUS protocol enhances state discrimination by partitioning the measurement outcome axis with two thresholds rather than a single one, creating three distinct regions: one corresponding to ground-state assignment, an excited-state region that triggers a conditional $X_\pi$ pulse, and an intermediate region of uncertainty. When a measurement outcome falls within the uncertain region, the protocol requires an additional measurement round, thereby increasing the likelihood of correct state identification.

A measurement performed on a qubit in the ground state yields a real outcome drawn from $\mathcal{N}(\mu_g, \sigma_g^2)$, while a qubit in the excited state produces an outcome from $\mathcal{N}(\mu_e, \sigma_e^2)$, with $\mu_g < \mu_e$. Two thresholds, $t_a < t_b$, partition the measurement axis into three regions, and the protocol acts based on the region in which the measurement falls:
\begin{equation}\label{eq:RUS}
r \;\longrightarrow\;
\begin{cases}
\text{assign } \lvert g\rangle, & r < t_a,\\[2pt]
\text{re-measure (no change)}, & t_a < r < t_b,\\[2pt]
\pi\text{-pulse, then re-measure}, & r > t_b.
\end{cases}
\end{equation}
Each qubit exits once a measurement outcome falls within $r < t_a$, and the final state assignment is made at the point of exit. These three regions correspond to six conditional probabilities: letting $\Phi_j(r) = \Phi\!\big((r-\mu_j)/\sigma_j\big)$ denote the cumulative distribution function of the Gaussian for state $j\in\{g,e\}$,
\begin{equation}
\label{eq:rates}
\begin{aligned}
(\alpha,\,\beta,\,\gamma) &= \big(\Phi_g(t_a),\; \Phi_g(t_b)-\Phi_g(t_a),\; 1-\Phi_g(t_b)\big), \\
(\delta,\,\epsilon,\,\eta) &= \big(\Phi_e(t_a),\; \Phi_e(t_b)-\Phi_e(t_a),\; 1-\Phi_e(t_b)\big),
\end{aligned}
\end{equation}
with $\alpha+\beta+\gamma=\delta+\epsilon+\eta=1$. Here $\alpha,\beta,\gamma$ are the probabilities that a ground-state qubit yields an outcome below $t_a$, between $t_a$ and $t_b$, and above $t_b$, respectively; $\delta,\epsilon,\eta$ are the corresponding probabilities for an excited-state qubit.

Let $G(i)$ and $E(i)$ denote the population fractions of qubits initialized in the ground and excited states, respectively, that remain in the protocol after $i$ rounds, with $G(0)=E(0)=0.5$. Thus, $G(i)$ and $E(i)$ represent the surviving ground- and excited-state populations after $i$ rounds, rather than the total ground- and excited-state populations at that stage since the total includes qubits that have exited the protocol. Applying the protocol rules from Eq.~\eqref{eq:RUS}, the evolution is described by the linear recursion
\begin{equation}
\label{eq:recursion}
\begin{pmatrix}G(i{+}1)\\ E(i{+}1)\end{pmatrix}
= M \begin{pmatrix}G(i)\\ E(i)\end{pmatrix},
\qquad
M = \begin{pmatrix}\beta & \eta\\ \gamma & \epsilon\end{pmatrix},
\end{equation}
so that $\bm{r}(i) = M^{i}\bm{r}_0$ with $\bm{r}_0 = (\tfrac{1}{2},\, \tfrac{1}{2})^{\mathsf{T}}$. The off-diagonal entries represent state conversion: $\eta$ is the probability that an excited state is flipped to ground, while $\gamma$ is the probability that a ground state is flipped to excited. The fractions of correct and incorrect exits occurring at round $i$ are
\begin{equation}
\label{eq:exitflux}
C(i) = \alpha G(i), \qquad W(i) = \delta E(i).
\end{equation}

The powers of $M$ over $i$ rounds follow from the Cayley--Hamilton theorem,
\begin{equation}
\label{eq:CH}
M^{i}=U_i M-D U_{i-1} I,
\qquad
U_i=\frac{\lambda_+^{i}-\lambda_-^{i}}{\lambda_+-\lambda_-},
\end{equation}
where $T$ and $D$ denote the trace and determinant of $M$, $\lambda_\pm$ are its eigenvalues, and $I$ is the identity matrix. The sequence $U_i$ satisfies $U_{i+1}=TU_i-DU_{i-1}$ with $U_0=0$, $U_1=1$. The eigenvalues are always real, since their discriminant simplifies to
\begin{equation}
\label{eq:disc}
T^2-4D=(\beta-\epsilon)^2+4\eta\gamma\ge 0,
\end{equation}
ensuring purely exponential population decay without oscillatory behavior, with $\lambda_\pm = \tfrac{1}{2}\!\left(\beta+\epsilon\pm\sqrt{(\beta-\epsilon)^2+4\eta\gamma}\right)$. This lets us compute the protocol's theoretical fidelity and convergence rate: the powers of $M$ determine the number of rounds required to reach a target fidelity, or conversely the expected fidelity after a fixed number of rounds $i$,
\begin{equation}
F_i = \frac{C(i)}{C(i)+W(i)} = \frac{\alpha G(i)}{\alpha G(i)+\delta E(i)},
\end{equation}
with $G(i)$, $E(i)$ given by Eq.~\eqref{eq:recursion}. The convergence rate and achievable fidelity are thus governed entirely by the placement of $t_a$ and $t_b$.

For a finite number of qubits $N$, the number of qubits that exit correctly by round $i$ follows a distribution with mean $NC(i)$:  The probability of a given qubit exiting correctly is $C(i)$, and of exiting incorrectly or remaining active is $1-C(i)$. Therefore, across $N$ independent runs, the correct exits are Bernoulli random variables with success probability $C(i)$, and a sum of $N$ independent Bernoulli random variables is, by definition, binomially distributed:
\begin{equation}
    \tilde{C}(i) = \mathrm{Binomial}(N,C(i)),
\end{equation}
with mean $NC(i)$ and variance $NC(i)(1-C(i))$. Similarly, the number of incorrect exits is $\tilde{W}(i)=\mathrm{Binomial}(N,W(i))$. Since $\tilde{C}(i)$ and $\tilde{W}(i)$ are two categories of a multinomial with $N$ trials (the third being the still-active population), $\mathrm{Cov}[\tilde{C}(i),\tilde{W}(i)] = -NC(i)W(i)$. Applying the multivariate delta method \cite{CasellaBerger2002} to $\tilde{F}(i)=\tilde{C}(i)/(\tilde{C}(i)+\tilde{W}(i))$ then gives, to leading order in $1/N$,
\begin{equation}
\mathrm{Var}[\tilde{F}(i)] \approx \frac{{F}(i)(1-{F}(i))}{N(C(i)+W(i))}.
\end{equation}
This expression neglects that the total protocol duration itself scales with $N$; if that duration exceeds the thermal-excitation timescale, the fidelity will decrease further.

We can compute how the RUS protocol's completion time scales with the number of qubits $N$. Let $A(i) = G(i)+E(i)$ denote the probability that a qubit remains active at round $i$; in terms of $M$, $A(i)=\bm{1}^{\mathsf{T}}M^i\bm{r}_0$, where $\bm{r}_0$ is the initial population vector and $\bm{1}^{\mathsf{T}}=[1,1]$. Since the eigenvalues of $M$ are known, this simplifies to
\begin{equation}
    A(i) = c_+ \lambda_+^i + c_- \lambda_-^i,
\end{equation}
where $c_+,c_-$ are set by the initial condition $A(0)=1$. As the eigenvalues of $M$ are less than 1, for large $i$ we can approximate $A(i)\approx c_+\lambda_+^i$. The probability that a given qubit has exited by round $i$ is $1-A(i)$, so for $N$ qubits, the probability that all $N$ have exited by round $i$ is $(1-A(i))^N$, and the probability that at least one qubit remains active beyond round $i$ is $A_N(i)=1-(1-A(i))^N$. The probability of being active exactly until round $i$ is then $P(i)=A_N(i)-A_N(i-1)$, giving the expected exit time
\begin{align}
    \mathbb{E}[T_N] &= \sum_{i=0}^\infty i\, P(i)
    = \sum_{i=0}^\infty A_N(i) \\
    &= \sum_{i=0}^{\infty} \left[1-(1-A(i))^N\right],
\end{align}
where the second line uses the standard identity $\mathbb{E}[T]=\sum_{i\ge0}P(T>i)$ for a nonnegative integer-valued random variable $T$.

\begin{figure}
\includegraphics[width=1\columnwidth]{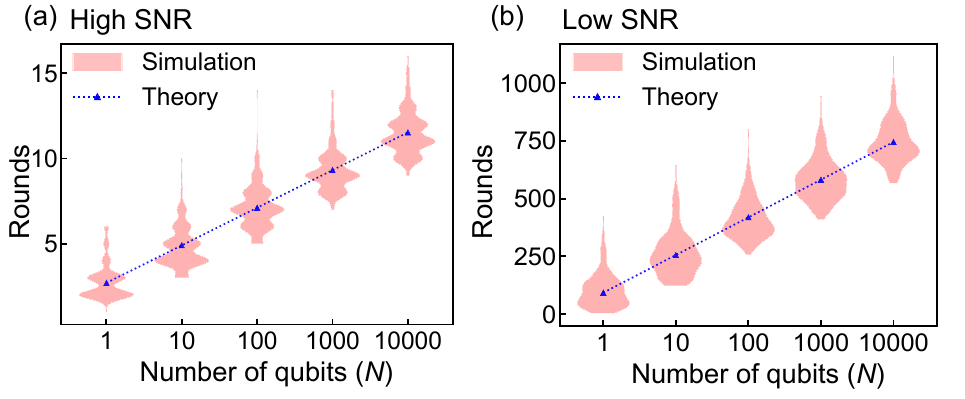}
\caption{
\textbf{Theory Prediction of exit time (Rounds) versus number of qubits.}  Expected number of rounds required for all $N$ qubits to exit the RUS protocol for (a) the high-SNR ($\mathrm{SNR}=2.11$) and (b) the low-SNR ($\mathrm{SNR}=1.05$) regime. Shaded violins show the distributions obtained from Monte Carlo simulations, while blue markers and dotted lines indicate the theoretical predictions. The high-SNR parameters correspond to those of $Q_2$ used in the main text, while the low-SNR case is obtained by halving the readout amplitude. For simplicity, all qubits are assumed to have identical readout parameters. In both regimes, the average exit time increases approximately logarithmically with the number of qubits, $T_N \propto \ln N$, with substantially longer exit times in the low-SNR regime. } 
\label{app:exit_time}
\end{figure}

For small $A(i)$, $(1-A(i))^N \sim e^{-Nc_+\lambda_+^i}$. Using the Euler--Maclaurin formula, we can write the discrete sum as the integral plus half the integrand's value at $i=0$ (dropping higher-order terms), $\mathbb{E}[T_N]\approx\frac{1}{2}(1-e^{-Nc_+})+\int_0^\infty\left(1-e^{-Nc_+\lambda_+^i}\right)di$. Defining $K=Nc_+$ and $m=\ln(1/\lambda_+)$, the integrand becomes $1-e^{-Ke^{-mi}}$; substituting $u=Ke^{-mi}$ gives $\mathbb{E}[T_N]=\frac12(1-e^{-K})+\frac{1}{m}\int_0^K\frac{1-e^{-u}}{u}\,du$. Integrating the remaining term by parts with $v=\ln u$, $dw=e^{-u}du$, the boundary term at $u=0$ vanishes since $u\ln u\to0$. This gives
\begin{equation}
\int_0^K\frac{1-e^{-u}}{u}\,du = (1-e^{-K})\ln(K) - \int_0^K e^{-u}\ln(u)\,du.
\end{equation}
For large $K$, the first term tends to $\ln K$, and the remaining integral tends to $\Gamma'(1)=\int_0^\infty e^{-u}\ln(u)\,du=-\gamma$ where $\Gamma(s)=\int_0^\infty u^{s-1}e^{-u}\,du$ being the Euler integral of the second kind and $\gamma$ the Euler--Mascheroni constant. The bracketed expression tends to $\ln K - (-\gamma) = \ln K+\gamma$. Plugging this back gives
\begin{align}
    \mathbb E[T_N] &\sim \frac{1}{2}+\frac{1}{m}[\ln K + \gamma ]\\
    &=\frac{1}{2}+\frac{\ln(Nc_+) + \gamma}{\ln(1/\lambda_+)}.
\end{align}
The expectation value of $T_N$ has no exact closed form, but for large $N$ and large $i$ it is well approximated by the expression above, which we verified against Monte Carlo simulation of the RUS protocol. This implies the exit time scales as $\ln N$. For a large number of qubits, most finish quickly but must idle until the last one exits, with the idling time growing logarithmically in $N$. Figure~\ref{app:exit_time}(a) and (b) show the expected exit time (rounds) increasing logarithmically with qubit number in the high-SNR and low-SNR regimes, respectively, consistent with the mean exit time from Monte Carlo simulation. We assume all qubits share the same Gaussian readout parameters as $Q_2$ in the experiments.

\section{Bayesian Inference for Qubit Readout with General Double-Gaussian Likelihoods} \label{App.theory calculation} 
We model the single-shot readout outcome in the $i$-th reset round as a real-valued quadrature variable $r_i$. The readout distributions conditioned on the qubit being in $\ket{g}$ or $\ket{e}$ are modeled as Gaussian likelihoods,
\begin{align}
p(r|g) &= \mathcal{N}(r;\mu_g,\sigma_g^2)
  = \frac{1}{\sqrt{2\pi}\,\sigma_g}
    \exp\!\left[-\frac{(r-\mu_g)^2}{2\sigma_g^2}\right],
  \label{eq:pg}\\[2pt]
p(r|e) &= \mathcal{N}(r;\mu_e,\sigma_e^2)
  = \frac{1}{\sqrt{2\pi}\,\sigma_e}
    \exp\!\left[-\frac{(r-\mu_e)^2}{2\sigma_e^2}\right],
  \label{eq:pe}
\end{align}
where $\mu_g,\sigma_g$ and $\mu_e,\sigma_e$ are obtained from calibration measurements of the ground- and excited-state readout distributions.

Before the $i$-th readout round, the prior ground-state probability is denoted
\begin{equation}
P_{i-1}(g) = \pi_{i-1}, \qquad P_{i-1}(e) = 1-\pi_{i-1}.
\label{eq:prior}
\end{equation}
After obtaining the measurement outcome $r_i$, the ground-state probability is updated according to Bayes' rule:
\begin{equation}
\pi_i \equiv P_i(g|r_i,r_{i-1},\ldots)
  = \frac{\pi_{i-1}\,p(r_i|g)}
         {\pi_{i-1}\,p(r_i|g)+(1-\pi_{i-1})\,p(r_i|e)}.
\label{eq:bayes}
\end{equation}
It is convenient to track the posterior log-odds, denoted
$\mathcal{L}_i$,
\begin{equation}
\mathcal{L}_i \equiv \ln\frac{\pi_i}{1-\pi_i}.
\label{eq:L-def}
\end{equation}
We define the single-shot likelihood ratio
\begin{equation}
R(r) \equiv \frac{p(r|e)}{p(r|g)},
\label{eq:R-def}
\end{equation}
and the associated log-likelihood ratio, denoted by lowercase $\ell$,
\begin{equation}
\ell(r) \equiv \ln\frac{p(r|g)}{p(r|e)} = -\ln R(r).
\label{eq:ell-def}
\end{equation}

Forming the ratio $\pi_i/(1-\pi_i)$ from Eq.~\eqref{eq:bayes} and its complement, the shared normalization cancels and the posterior odds factorize as
\begin{equation}
\frac{\pi_i}{1-\pi_i}
  = \frac{\pi_{i-1}}{1-\pi_{i-1}}\cdot\frac{p(r_i|g)}{p(r_i|e)}.
\label{eq:odds-factorize}
\end{equation}
Taking the logarithm of Eq.~\eqref{eq:odds-factorize} and using Eqs.~\eqref{eq:L-def} and \eqref{eq:ell-def} gives the additive recursion
\begin{equation}
\mathcal{L}_i = \mathcal{L}_{i-1} + \ell(r_i),
\label{eq:L-recursion}
\end{equation}
which, upon recursion \cite{Wald1945}, gives
\begin{equation}
\mathcal{L}_i = \mathcal{L}_0 + \sum_{k=1}^{i}\ell(r_k),
\qquad
\mathcal{L}_0 = \ln\frac{\pi_0}{1-\pi_0}.
\label{eq:L-sum}
\end{equation}
For an unbiased initial prior, $\mathcal{L}_0=0$. The posterior probability is recovered from $\mathcal{L}_i$ via the sigmoid
\begin{equation}
\pi_i = \frac{1}{1+e^{-\mathcal{L}_i}}.
\label{eq:sigmoid}
\end{equation}
The conditional $X_\pi$ pulse is applied whenever $\mathcal{L}_i<0$, i.e., whenever the accumulated evidence favors $\ket{e}$ over
$\ket{g}$.

Substituting Eqs.~\eqref{eq:pg}--\eqref{eq:pe} into Eq.~\eqref{eq:R-def}, the likelihood ratio takes the form
\begin{align}
R(r) &= \frac{\sigma_g}{\sigma_e}
  \exp\!\left[-\frac{(r-\mu_e)^2}{2\sigma_e^2}
              +\frac{(r-\mu_g)^2}{2\sigma_g^2}\right]\\
  &= D\exp\!\left(Ar^2+Br+C\right),
\label{eq:R-gaussian}
\end{align}
where
\begin{align}
A &= \frac{1}{2\sigma_g^2}-\frac{1}{2\sigma_e^2}, &
B &= -\frac{\mu_g}{\sigma_g^2}+\frac{\mu_e}{\sigma_e^2},
  \label{eq:AB}\\
C &= \frac{\mu_g^2}{2\sigma_g^2}-\frac{\mu_e^2}{2\sigma_e^2}, &
D &= \frac{\sigma_g}{\sigma_e}.
  \label{eq:CD}
\end{align}

Correspondingly, from Eq.~\eqref{eq:ell-def},
\begin{equation}
\ell(r) = -Ar^2 - Br - C - \ln D.
\label{eq:ell-closed}
\end{equation}
For $\sigma_g=\sigma_e\equiv\sigma$, Eq.~\eqref{eq:ell-closed} reduces to the linear form $\ell(r) = (\mu_g-\mu_e)(r-\bar\mu)/\sigma^2$, with $\bar\mu \equiv (\mu_g+\mu_e)/2$.

The threshold $t_i$ for round $i$ is defined as the value of $r$ at which the two prior-weighted likelihoods are equal,
\begin{equation}
\pi_{i-1}\,p(t_i|g) = (1-\pi_{i-1})\,p(t_i|e),
\label{eq:threshold-def}
\end{equation}
equivalent to the condition $\ell(t_i) = -\mathcal{L}_{i-1}$. Substituting Eq.~\eqref{eq:ell-closed} gives the quadratic
\begin{equation*}
A t_i^2 + B t_i + E_{i-1} = 0,
\qquad
E_{i-1} \equiv C + \ln D - L_{i-1}.
\end{equation*}
with roots
\begin{equation}
t_i = \frac{-B \pm \sqrt{B^2-4 A E_{i-1}}}{2A}.
\label{eq:roots}
\end{equation}

The maximum prior probability $\pi_{\max}$ is reached when the two threshold solutions merge, corresponding to a vanishing discriminant, $B^2-4AE=0$. This gives
\begin{equation}
\ln\left(\frac{1-\pi_{\max}}{\pi_{\max}}\right)
=
-\ln\left(\frac{\sigma_g}{\sigma_e}\right)
-
\frac{(\mu_g-\mu_e)^2}
{2(\sigma_g^2-\sigma_e^2)}.
\end{equation}

Therefore,
\begin{equation}
\pi_{\max}
=
1-
\frac{1}{
1+
\dfrac{\sigma_g}{\sigma_e}
\exp\left[
\frac{(\mu_g-\mu_e)^2}
{2(\sigma_g^2-\sigma_e^2)}
\right]
}.
\end{equation}

The physically relevant root is the one lying between $\mu_g$ and $\mu_e$. Starting from an unbiased prior $\pi_0=1/2$, the first threshold is obtained from the calibrated Gaussian distributions; after each measurement outcome $r_i$, the posterior $\pi_i$ is evaluated, an $X_\pi$ pulse is applied if it favors $\ket{e}$, the posterior is updated as $\pi_i\rightarrow1-\pi_i$, and Eq.~\eqref{eq:roots} is recomputed for the next round using this updated prior. In this way, the threshold adapts dynamically to the accumulated measurement information.

\section{Device Setup} 
\label{app:device}
Our experiments are performed on a five-qubit superconducting processor based on fixed-frequency transmons and tunable couplers. Table~\ref{tab:qubit_parameters} lists the main characteristics of the device. For each qubit $Q_i$, we report its transition frequency $\omega_\mathrm{q}/2\pi$, anharmonicity $\eta/2\pi$, relaxation time $T_1$, Ramsey coherence time $T_2^*$, and Hahn-echo coherence time $T_{2,\mathrm{echo}}$. State readout is provided by a dedicated resonator coupled dispersively to each qubit. The corresponding resonator parameters include the resonance frequency $\omega_{\mathrm r}/2\pi$, linewidth $\kappa_{\mathrm r}/2\pi$, and dispersive shift $\chi/2\pi$.

\begin{table}[htbp]
    \centering
    \caption{Qubit parameters.}
    \label{tab:qubit_parameters}
    \renewcommand{\arraystretch}{1.3}
    \setlength{\tabcolsep}{7pt}
    \begin{tabular}{c|ccccc}
        \hline\hline
         & $Q_1$ & $Q_2$ & $Q_3$ & $Q_4$ & $Q_5$ \\
        \hline
        $\omega_\mathrm{q}/2\pi$ (GHz)
        &4.244 &4.076 &3.996 &3.938 &3.837 \\

        $\eta/2\pi$ (MHz)
        &$-145$ &$-144$ &$-146$ &$-147$ &$-148$ \\

        $T_1$ ($\mu$s)
        &25 &59 &60 &62 &19 \\

        $T_2^*$ ($\mu$s)
        &24 &27 &30 &27 &30 \\

        $T_{2,\mathrm{echo}}$ ($\mu$s)
        &38 &105 &85 &92 & 38 \\

        $\omega_\mathrm{r}/2\pi$ (GHz)
        &7.726 &7.644 &7.574 &7.506 &7.439 \\

        $\kappa_\mathrm{r}/2\pi$ (MHz)
        &0.566 &1.067 &2.113 &1.511 &1.387 \\

        $\chi/2\pi$ (MHz)
        &0.118 &0.133 &0.108 &0.115 &0.127 \\
        \hline\hline
    \end{tabular}
\end{table}

\section{Real-Time Control Sequence and Implementation} 
The adaptive-reset protocol is implemented using the real-time processing and conditional-control capabilities of the FPGA in the OPX$+$.  To implement the Bayesian update and adaptive-threshold calculation discussed in App.~\ref{App.theory calculation} while reducing the computational overhead of real-time FPGA processing, we perform the Bayesian update in the logarithmic domain. We introduce the transformed state variable
\begin{equation}
s_i=\frac{\ln [(1-\pi_i)/\pi_i]}{A}.
\end{equation}
For each round, the OPX$+$ acquires the integrated in-phase quadrature $r_i$ and updates the state variable according to
\begin{equation}
s_i \rightarrow s_i + (r_i-R_1)(r_i-R_2)+H,
\end{equation}
where
\begin{equation}
R_{1,2}
=
\frac{-B\pm\sqrt{B^2-4AC}}{2A},
\qquad
H=\frac{\ln D}{A}.
\end{equation}
Here, $R_1$ and $R_2$ are the two roots of $Ar^2+Br+C=0$. The measured value is then compared with the current adaptive threshold $t_i$. If $r_i>t_i$, the OPX$+$ conditionally applies an $X_\pi$ pulse and updates $s\rightarrow -s$, reflecting the interchange of the ground- and excited-state populations, i.e., $\pi_i \rightarrow 1-\pi_i$. Finally, the threshold for the next round is calculated in real time as
\begin{equation}
t_{i+1}=-F+s_{\mathrm{root}}\sqrt{G-(s+H)},
\end{equation}
where
\begin{equation}
F=\frac{B}{2A},
\qquad
G=F^2-\frac{C}{A},
\end{equation}
and $s_{\mathrm{root}}\in\{+1,-1\}$ denotes the sign of the physical root selected during initialization. Specifically, $s_{\mathrm{root}}$ is chosen such that the initial threshold
\begin{equation}
t_{\mathrm{init}}
=
-F+s_{\mathrm{root}}\sqrt{G-H}
\end{equation}
lies between the centers of the ground- and excited-state readout distributions, and the same procedure is repeated for subsequent reset rounds. 

\label{app:timing}
\begin{figure}
\includegraphics[width=\columnwidth]{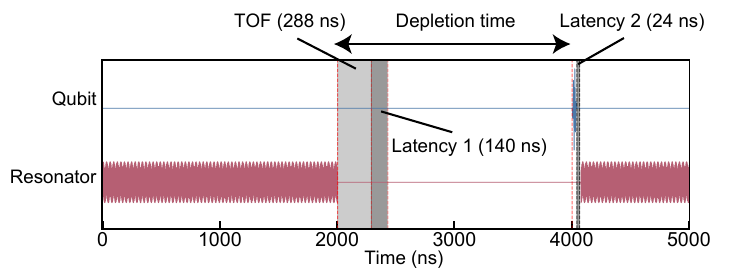}
\caption{
\textbf{Timing diagram.} Real-time pulse sequence for one round of the adaptive active-reset protocol implemented on Q$_2$ and its corresponding readout resonator. Each round begins with a $2~\mu\mathrm{s}$ measurement pulse, followed by a $288~\mathrm{ns}$ time of flight (TOF) and a $140~\mathrm{ns}$ processing latency (Latency 1) for updating the prior. Both the TOF and Latency 1 are accommodated within the $2~\mu\mathrm{s}$ resonator-depletion window. Based on the measurement outcome and the updated prior, a conditional $X_\pi$ pulse is then applied to the qubit. Finally, an additional $24~\mathrm{ns}$ processing latency (Latency 2) is required to calculate the threshold for the subsequent round before the next measurement begins.
}
\label{fig:fig7}
\end{figure}
The real-time pulse sequence used to implement the above protocol is shown in Fig.~\ref{fig:fig7}. The adaptive-reset protocol is demonstrated on Q$_2$, with both the measurement duration and the depletion time of the resonator set to $2~\mu\mathrm{s}$. The real-time Bayesian update and threshold calculation are performed within the sequence as described above. In the timing diagram, we illustrate the branch in which the measurement outcome triggers the conditional $X_\pi$ pulse.

\section{Distribution of the Adaptive Threshold}
\label{app:threshold-dist}
The threshold is a deterministic function of the accumulated log-odds $\mathcal{L}_{i-1}$, which in turn depends on the random measurement record $\{r_1,\ldots,r_{i-1}\}$. We first treat the case of equal variances, $\sigma_g=\sigma_e\equiv\sigma$ and $\mu_g\neq\mu_e$, for which the threshold at round $i$ follows directly from Eq.~\eqref{eq:roots},
\begin{equation}
t_i = \bar\mu + \frac{\sigma^2}{\mu_e-\mu_g}\,\mathcal{L}_{i-1},
\qquad
\bar\mu \equiv \frac{\mu_g+\mu_e}{2}.
\label{eq:tE1}
\end{equation}
For an unbiased initial prior, $\mathcal{L}_0=0$ and
\begin{equation}
\mathcal{L}_{i-1} = \sum_{k=1}^{i-1}\ell(r_k).
\label{eq:tE1b}
\end{equation}
The expectation value of the threshold, given that the true state is
$g$, is
\begin{equation}
    \bar{t}_i = \bar{\mu}+\frac{\sigma^2}{\mu_e-\mu_g}\langle \mathcal{L}_{i-1}\rangle.
\end{equation}
Each term in Eq.~\eqref{eq:tE1b} has an expectation value given by the
Kullback-Leibler divergence,
\begin{equation}
\mathbb{E}[\ell|g] = \int p(r|g)\ln\frac{p(r|g)}{p(r|e)}\,dr
  \equiv \mathcal{D},
\label{eq:tE3}
\end{equation}
which for two equal-variance Gaussians evaluates to $\mathcal{D}=\Delta\mu^2/(2\sigma^2)$, with $\mathrm{Var}[\ell|g]=2\mathcal{D}$, where $\Delta\mu \equiv \mu_g-\mu_e$. The expectation value of the threshold after $i-1$ rounds, given that the true state is $g$ throughout, is therefore
\begin{equation}
\bar t_{i}\big|_{g} = \bar\mu + \frac{\sigma^2}{\mu_e-\mu_g}(i-1)\mathcal{D}
  = \bar\mu - \frac{\Delta\mu}{2}(i-1).
\label{eq:tE4}
\end{equation}
Thus the threshold moves, on average, away from $\bar\mu$ toward $\mu_e$ when the true state is $g$ (given $\mu_e>\mu_g$ in our convention), becoming progressively more conservative. If the true state is instead $e$, $\mathbb{E}[\ell|e]=-\mathcal{D}$, and the threshold moves on average toward $\mu_g$.

From Eq.~\eqref{eq:ell-closed} with $A=0$, the log-likelihood per step is linear in $r$ and therefore Gaussian; the log-odds $\mathcal{L}_{i-1}$, a sum of independent Gaussian increments, is itself Gaussian with the mean and variance given above:
\begin{equation}
t_i\big|_{g} \sim \mathcal{N}\!\left(\bar\mu - \frac{\Delta\mu}{2}(i-1),\,
  \sigma^2(i-1)\right),
\qquad t_i\geq\bar\mu,
\label{eq:tE4b}
\end{equation}
truncated and renormalized on $[\bar\mu,\infty)$. This description assumes the true state is $g$ throughout the run. In the adaptive protocol discussed in the paper, the $X_\pi$ pulse correlates the true state with the belief history, so this Gaussian form holds only asymptotically. In the initial rounds, where the true state may still be $e$, the marginal threshold distribution is a mixture of two folded Gaussians,
\begin{align*}
t_{i}\Big|_{\text{marginal}} &\sim
  \mathcal{N}\!\left(\bar\mu-\frac{\Delta\mu}{2}(i-1),\sigma^2(i-1)\right)\\
  &+ \mathcal{N}\!\left(\bar\mu+\frac{\Delta\mu}{2}(i-1),\sigma^2(i-1)\right),
\label{eq:tE5}
\end{align*}
with $t_i\geq\bar\mu$; the first term comes from true-$g$ trajectories and the second from true-$e$ trajectories. Early in the protocol this produces a lifted tail on the low-$t$ side of the distribution, which decays as the true-state-$e$ contribution moves away from $\bar\mu$ and its weight in the mixture diminishes.

For $\sigma_g\neq\sigma_e$, the log-likelihood per step is quadratic in $r$ [Eq.~\eqref{eq:ell-closed}], and the threshold [Eq.~\eqref{eq:roots}] is no longer linear in $\mathcal{L}_{i-1}$. The expectation value of $\ell$ given the true state is $g$ is now
\begin{equation}
\mathbb{E}[\ell|g] = \ln\frac{\sigma_e}{\sigma_g}
  + \frac{\sigma_g^2+\Delta\mu^2}{2\sigma_e^2} - \frac{1}{2},
\label{eq:tE6}
\end{equation}
with the $e$-conditioned expectation obtained by exchanging $g\leftrightarrow e$; the two KL divergences are in general unequal. The single-shot increment is a scaled, shifted square of $r_i$,
\begin{equation}
\ell(r_i) = -A\!\left(r_i+\frac{B}{2A}\right)^{\!2} + \frac{B^2}{4A} - C - \ln D.
\label{eq:tE7}
\end{equation}
If the qubit is in state $s\in\{g,e\}$, with $r_i\sim \mathcal{N}(\mu_s,\sigma_s^2)$, then $(r_i+B/2A)/\sigma_s$ is a unit Gaussian displaced from zero, and its square is a noncentral $\chi^2$ random variable,
\begin{equation}
\ell = -A\sigma_s^2\,\chi^2(\lambda_s) + \frac{B^2}{4A} - C - \ln D,
\qquad
\lambda_s = \left(\frac{\mu_s+B/2A}{\sigma_s}\right)^{\!2}.
\label{eq:tE8}
\end{equation}
The mean of this distribution reproduces the KL divergence, Eq.~\eqref{eq:tE6}, as required. In an adaptive protocol with $X_\pi$ pulses, the per-step increment can be drawn from either state's distribution depending on the run's history, so this closed form only describes $\ell(r_i)$ conditioned on a fixed true state.

Because the true state at a given round is not known in advance, we construct the belief distribution directly. Let $f_g(\mathcal{L},i)$ and $f_e(\mathcal{L},i)$ denote the population-weighted densities of $\mathcal{L}_i$ for trajectories whose true state is $g$ or $e$, respectively. A measurement displaces the belief by convolution with the single-shot increment density $w_s(\ell)$ [Eq.~\eqref{eq:tE8}]:
\begin{equation}
\tilde f_s(\mathcal{L}) = \int f_s(\mathcal{L}-\ell,i)\,w_s(\ell)\,d\ell.
\label{eq:tE9}
\end{equation}
Whenever $\mathcal{L}<0$, an $X_\pi$ pulse is applied, mapping $\mathcal{L}\to-\mathcal{L}$ and exchanging the true-state label. The post-pulse densities for round $i+1$ are therefore
\begin{align}
f_g(\mathcal{L},i+1) &= \tilde f_g(\mathcal{L}) + \tilde f_e(-\mathcal{L}),
\label{eq:tE10a}\\
f_e(\mathcal{L},i+1) &= \tilde f_e(\mathcal{L}) + \tilde f_g(-\mathcal{L}).
\label{eq:tE10b}
\end{align}

Since $t_i$ is related to $\mathcal{L}_{i-1}$ through the threshold condition $\mathcal{L}_{i-1}=-\ell(t_i)$, the event that $\mathcal{L}_{i-1}$ falls in a small interval $[\mathcal{L},\mathcal{L}+d\mathcal{L}]$ corresponds to the event that $t_i$ falls in the corresponding interval $[t_i,t_i+dt_i]$. Equating the probability assigned to this event under each description,
\begin{equation}
f(\mathcal{L},i-1)\,d\mathcal{L} = p(t_i)\,dt_i,
\label{eq:tE10c}
\end{equation}
and substituting $\mathcal{L}=-\ell(t_i)$ together with $d\mathcal{L}=-\ell'(t_i)\,dt_i$ gives the threshold density in terms of the belief density,
\begin{equation}
p(t_i) = f(-\ell(t_i),i-1)\,\left|\frac{d\ell(t_i)}{dt_i}\right|.
\label{eq:tE10d}
\end{equation}
Writing the total belief density as the sum of its two population-weighted, true-state-conditioned pieces, $f(\mathcal{L},i-1)=f_g(\mathcal{L},i-1)+f_e(\mathcal{L},i-1)$, gives the full threshold distribution
\begin{equation}
p(t_i) = \left[f_g(-\ell(t_i),i-1) + f_e(-\ell(t_i),i-1)\right]
  \left|\frac{d\ell(t_i)}{dt_i}\right|.
\label{eq:tE11}
\end{equation}
We checked the empirical results against expected trajectories of the above analysis.
\begin{figure}[h]
\includegraphics[width=\columnwidth]{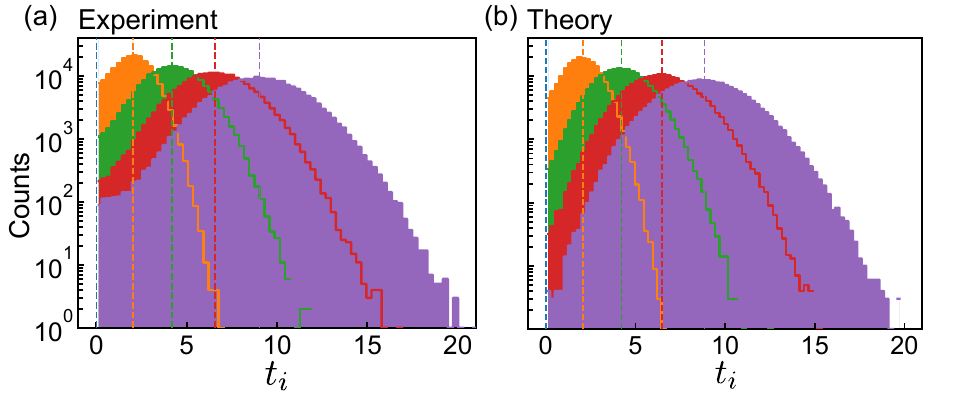}
\caption{
\textbf{Threshold distribution.} (a) Distributions of the reset threshold $t_i$ at each round. Vertical dashed lines indicate the mean threshold values. (b) Distributions of the reset threshold calculated by KL divergence. 
}
\label{fig:fig8}
\end{figure}
Figure~\ref{fig:fig8} compares the experimental and theoretical threshold distributions at each measurement round. The theoretical distributions are obtained numerically by discretizing the belief axis with a bin size of $\Delta=0.02$ and restricting the range to $|\mathcal{L}|<200$, chosen so that, for the experimental parameters used, the values remain within the real root of the thresholds. The overall agreement between experiment and theory is good; the remaining deviations may arise from $T_1$ relaxation and imperfections in modeling the readout distributions as two Gaussians.

\section{Entropy and Information Gain} \label{app:entropy}
\begin{figure}
\includegraphics[width=\columnwidth]{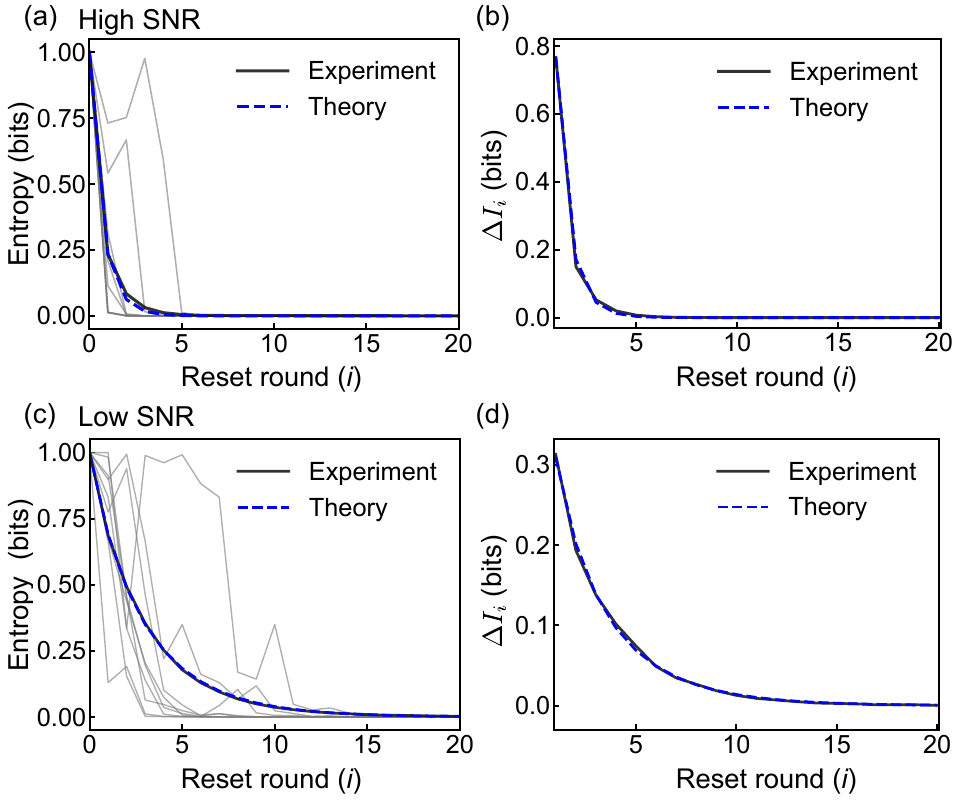}
\caption{
\textbf{Entropy and marginal information gain.}  (a,b) Binary entropy $H$ and marginal information gain after another round of reset $\Delta I_i$ versus reset round for $\mathrm{SNR}=2.11$. Gray curves show individual trajectories, the blue curve shows the experimental mean, and the black curve shows the prediction obtained from the KL divergence. (c,d) Same as (a) and (b), but for $\mathrm{SNR}=1.05$. }

\label{fig:fig9}
\end{figure}
The adaptive measurement-based reset can be viewed as successive entropy removal from the observer's belief about the qubit state. If the posterior after $i$ rounds is $\pi_i$, the binary entropy of the belief is
\begin{equation}
H(\pi_i) = -\pi_i\log_2\pi_i - (1-\pi_i)\log_2(1-\pi_i).
\label{eq:F1}
\end{equation}
Substituting the sigmoid relation between $\pi_i$ and the log-odds $\mathcal{L}_i$ [Eq.~\eqref{eq:sigmoid}] gives the entropy directly as a function of $\mathcal{L}_i$,
\begin{equation}
H(\mathcal{L}_i) = \frac{1}{\ln 2}\left[\ln\!\left(1+e^{-\mathcal{L}_i}\right)
  + \frac{\mathcal{L}_i}{1+e^{\mathcal{L}_i}}\right].
\label{eq:F2}
\end{equation}
For an unbiased prior, $\mathcal{L}_0=0$ and $H(\mathcal{L}_0)=1$; the entropy falls to zero once the state is known with certainty ($\mathcal{L}_i\to\pm\infty$). Individual trajectories may transiently increase $H$, but the ensemble average decreases monotonically toward zero.

We restrict the analysis to $\sigma_g=\sigma_e\equiv\sigma$, since in our experiment the two readout distributions differ predominantly in mean rather than variance; the general unequal-variance case can be treated via the noncentral-$\chi^2$ formalism of App.~\ref{app:threshold-dist}. 

Using $\Delta\mu\equiv\mu_g-\mu_e$ and $\bar\mu\equiv(\mu_g+\mu_e)/2$ as defined in App.~\ref{app:threshold-dist}, the single-shot log-likelihood ratio reduces to the linear form $\ell(r_i)=(\Delta\mu/\sigma^2)(r_i-\bar\mu)$. Its expectation value under the true state $g$ is the Kullback-Leibler divergence $\mathbb{E}[\ell|g]=\mathcal{D}=\Delta\mu^2/(2\sigma^2)$ [Eq.~\eqref{eq:tE3}], with variance $\mathrm{Var}[\ell|g]=2\mathcal{D}$. Consequently $\mathcal{L}_i$, a sum of $i$ such increments, is Gaussian with mean $i\mathcal{D}$ and variance $2i\mathcal{D}$ (App.~\ref{app:threshold-dist}). This can be used to compute the expectation value of the entropy of
belief,
\begin{equation}
\mathbb{E}[H_i] = \int H(\mathcal{L})\,\mathcal{N}(\mathcal{L}; i\mathcal{D}, 2i\mathcal{D})\,d\mathcal{L}.
\label{eq:F4}
\end{equation}
Although the true state is a priori unknown and could equally be $g$ or $e$, Eq.~\eqref{eq:F4} already gives the exact marginal expectation: since $H(\mathcal{L})=H(-\mathcal{L})$, the mirror contribution from the opposite true state integrates to the identical value.

The total information gained after $i$ rounds, relative to starting with complete uncertainty, is
\begin{equation}
I_i = 1 - \mathbb{E}[H_i].
\label{eq:F4b}
\end{equation}
For an unbiased prior this is also the mutual information between the true qubit state and the measurement record. The marginal information gain from the $i$-th reset round is
\begin{equation}
\Delta I_i = I_i - I_{i-1} = \mathbb{E}[H_{i-1}] - \mathbb{E}[H_i].
\label{eq:F5}
\end{equation}
As expected, the information gained per round is largest in the first few iterations, when uncertainty is highest, and diminishes as the posterior accumulates near the boundaries. Rather than targeting a fixed fidelity, the protocol can instead be designed to stop once the expected information gain from an additional measurement falls below a threshold $\epsilon$, beyond which further measurements provide negligible information. We define this information-theoretic stopping time as
\begin{equation}
i_\epsilon^{*} = \min\{i : \Delta I_i < \epsilon\}.
\label{eq:F6}
\end{equation}
This stopping point is a property of the measurement process alone and is independent of the number of qubits being reset: all $N$ qubits can be run for the same number of rounds, set by where the marginal information gain becomes negligible. Increasing $N$ changes the simultaneous-reset fidelity but not the stopping point itself.  In contrast, repeat-until-success protocols terminate only once every qubit has individually exited the uncertainty region, giving a completion time that grows logarithmically with system size.

Figure~\ref{fig:fig9}(a) and (c) compare the experimental and theoretical entropy at each measurement round for the high- and low-SNR cases (SNR~$=2.11$ and $1.05$, respectively). In both regimes the entropy decreases as successive measurements accumulate information about the qubit state, with substantially faster decay at high SNR. As shown in Figs.~\ref{fig:fig9}(b) and (d), the marginal information gain $\Delta I_i$ decreases rapidly with round number; once it becomes negligible, additional measurement rounds provide essentially no further information. This behavior is consistent with the fidelity curves in Fig.~\ref{fig:fig3}(d), where the fidelity saturates after approximately $6$ rounds.

\end{document}